\documentclass[twocolumn,notitlepage,showkeys,showpacs,prl]{revtex4-1}
\usepackage{graphicx}
\usepackage{dcolumn}
\usepackage{booktabs,bm,color,braket,amsmath}
\usepackage{txfonts}
\usepackage{ulem}
\usepackage{color}
\usepackage[dvipsnames]{xcolor}
\usepackage[colorlinks,linkcolor=blue,citecolor=blue,urlcolor=blue]{hyperref}

\begin{document}
\title{Ubiquitous Topological States of Phonons in Solids: Silicon as a Model Material}
\author{Yizhou \surname{Liu}$^{1,3}$}
\author{Nianlong \surname{Zou}$^{1}$}
\author{Sibo \surname{Zhao}$^{1}$}
\author{Xiaobin \surname{Chen}$^{4,5,6}$}
\email{chenxiaobin@hit.edu.cn}
\author{Yong \surname{Xu}$^{1,2,7,8}$}
\email{yongxu@mail.tsinghua.edu.cn}
\author{Wenhui Duan$^{1,2,7,9}$}

\affiliation{
$^1$State Key Laboratory of Low Dimensional Quantum Physics, Department of Physics, Tsinghua University, Beijing 100084, China.
$^2$Tencent, Shenzhen, Guangdong 518057, China.
$^3$Department of Condensed Matter Physics, Weizmann Institute of Science, Rehovot 76100, Israel.
$^4$School of Science, Harbin Institute of Technology, Shenzhen 518055, China.
$^5$State Key Laboratory on Tunable laser Technology and Ministry of Industry and Information Technology Key Lab of Micro-Nano Optoelectronic Information System, Harbin Institute of Technology, Shenzhen 518055, China.
$^6$Collaborative Innovation Center of Extreme Optics, Shanxi University, Taiyuan 030006, China.
$^7$Frontier Science Center for Quantum Information, Beijing 100084, China.
$^8$RIKEN Center for Emergent Matter Science (CEMS), Wako, Saitama 351-0198, Japan.
$^9$Institute for Advanced Study, Tsinghua University, Beijing 100084, China.}

\begin{abstract}
Research on topological physics of phonons has attracted enormous interest but demands appropriate model materials. Our {\it ab initio} calculations identify silicon as an ideal candidate material containing extraordinarily rich topological phonon states. In silicon, we identify various topological nodal lines protected by glide mirror or mirror symmetries and characterized by quantized Berry phase $\pi$, which gives drumhead surface states observable from any surface orientations. Remarkably, a novel type of topological nexus phonon is discovered, which is featured by double Fermi-arc-like surface states and distinguished from Weyl phonons by requiring neither inversion nor time-reversal symmetry breaking. Versatile topological states can be created from the nexus phonons, such as Hopf nodal link by strain. Furthermore, we generalize the symmetry analysis to other centrosymmetric systems and find numerous candidate materials, demonstrating the ubiquitous existence of topological phonons in solids. These findings open up new opportunities for studying topological phonons in realistic materials and their influence on surface physics.
\end{abstract}

\maketitle

\textit{Introduction.}---Exploring novel quantum degrees of freedom to control phonons is of crucial importance to fundamental science and practical applications. Recent discoveries of Berry-phase and topological physics shed new lights on this subject, leading to an emerging field of {\it topological phononics} \cite{li2012RMP,Huber2016NatPhys, Susstrunk2016PNAS, LiuYZ2017NSR, MaGC2019NatRevPhys, LiuYZ2020AFM}. So far various kinds of topological phases of phonons have been proposed by theoretical models, including topological gapped and gapless phases in both two dimensions (2D) and 3D \cite{Prodan2009PRL, ZhangLF2010PRL, WangYT2015NJP, WangP2015PRL, Khanikaev2015NatCommun, LiuYZ2017PRB, Kane2014NatPhys, Susstrunk2015Science, LiuYZ2017PRL, XiaoM2015NatPhys_Synthetic, LiF2018NatPhys, ZhangTT2018PRL, MiaoH2018PRL, LiJX2018PRB, XiongZ2018PRB, LiuYZ2019Research, XieBY2019PRL, YangYH2019NatPhys, LiuQB2019JPCL, XiaBW2019PRL, ZhangTT2019PRL, WangR2020PRL, LiuQB2020npjComputMater, LiuPF2021PRB, ZhangTT2020PRR, ZhangTT2020PRB, LiuQB2021PRB, LiJX2021NatCommun, Chen2021PRL}; exotic topological and Berry-phase effects have also been proposed, which could find promising applications in thermal diodes \cite{LiuYZ2017PRB}, transistors \cite{LiuYZ2019Research}, antennas \cite{ZhangZW2018AM}, etc. However, very little experimental progress has been achieved in this field, mainly due to the lack of suitable realistic candidate materials.

Topological states are usually associated with nontrivial Berry phase, which can be defined by integrating either Berry connection along a closed 1D loop or Berry curvature on a 2D closed surface where bands are gapped. However, combined inversion-time-reversal ($\mathcal{PT}$) symmetry ensures that the Berry curvature vanishes everywhere except at band-degenerate nodes \cite{LiuYZ2020AFM}. Thus two general routes have been tried to search for phononic topological materials. One route requires breaking of $\mathcal{PT}$ symmetry and the other one requires protection of band degeneracy. For the first route (\textit{i.e.} breaking $\mathcal{PT}$ symmetry) there exist two scenarios: to break time-reversal symmetry (TRS) $\mathcal{T}$ or to break the inversion symmetry $\mathcal{P}$. While breaking TRS can lead to novel topological states such as the quantum-anomalous-Hall-like states \cite{Prodan2009PRL, ZhangLF2010PRL, WangYT2015NJP, WangP2015PRL, Khanikaev2015NatCommun, LiuYZ2017PRB}, it is difficult to realize strong TRS breaking effects of phonons in solids. By contrast, the $\mathcal{P}$ symmetry is naturally broken in noncentrosymmetric materials, where various kinds of Weyl phonons including double Weyl \cite{ZhangTT2018PRL, MiaoH2018PRL, LiJX2018PRB}, mixed single-double Weyl \cite{WangR2020PRL, LiuQB2020npjComputMater, LiuPF2021PRB}, quadruple Weyl \cite{ZhangTT2020PRB}, and charge-four Weyl \cite{LiuQB2021PRB} were revealed recently. Along the second route, band degeneracies protected by crystalline symmetries are generally demanded to induce 2D topological states such as the quantum-spin-Hall-like states \cite{Susstrunk2015Science, LiuYZ2017PRL}. The relevant crystalline symmetry, however, usually gets broken at the 1D boundary, which destroys the topological protection of edge states \cite{LiuYZ2017PRL}. Such a problem can be avoided in 3D materials. For instance, rotational and mirror symmetries could be preserved on some 2D surfaces \cite{LiuYZ2019Research}. Recent works proposed a few candidate materials of topological nodal-line phonons \cite{XiongZ2018PRB, LiuYZ2019Research}. These previous works mainly focused on noncentrosymmetric materials. The large class of centrosymmetric materials which may host topological phonon states, however, remain largely unexplored~\cite{LiJX2021NatCommun, Chen2021PRL}.

Among centrosymmetric materials, silicon (Si) is one of the most important materials for semiconductor technology and industry. Exploration of its surface phonons dates back to 1971, when Ibach discovered a dipole-active optical surface mode of about 55 meV on a clean Si(111) surface~\cite{Ibach1971PRL}. The surface mode was confirmed to be within 55-60 meV by others \cite{DiNardo1986PRB, Harten1988PRB, Pennino1989PRB}. The mysterious surface phonon has attracted intensive experimental and theoretical interest \cite{DiNardo1986PRB, Harten1988PRB, Pennino1989PRB, Alerhand1985PRL, Miglio1989PRL, Ancilotto1990PRL} but not been fully understood. Whether there exists a topological origin or not was never considered.

In this Letter, we find in diamond-structure Si plenty of topological nodal-line phonons, which have quantized Berry phase $\pi$ and are protected by glide-mirror or mirror symmetries, leading to drumhead surface phonon modes detectable for arbitrary surface orientations. Particularly, we predict a new kind of topological nexus phonon located at $\sim$55 meV in Si coinciding with the energy of surface phonon mode from Ibach's experiment \cite{Ibach1971PRL}. The nexus phonons can be viewed as combinations of pairs of Weyl phonons with opposite chiralities and protected against mutual annihilation by skew rotation symmetry, giving double Fermi-arc-like surface modes. The nexus phonons possess highly tunable topological features, which could evolve into pairs of Weyl phonons with opposite chiralities under TRS breaking and into interlocking topological Hopf nodal link under external strains. Moreover, we further inspect the symmetry of other centrosymmetric space groups and identify numerous candidate materials, revealing the ubiquitous existence of topological phonons in solids.

\begin{figure}
\centering
\includegraphics[width=\linewidth]{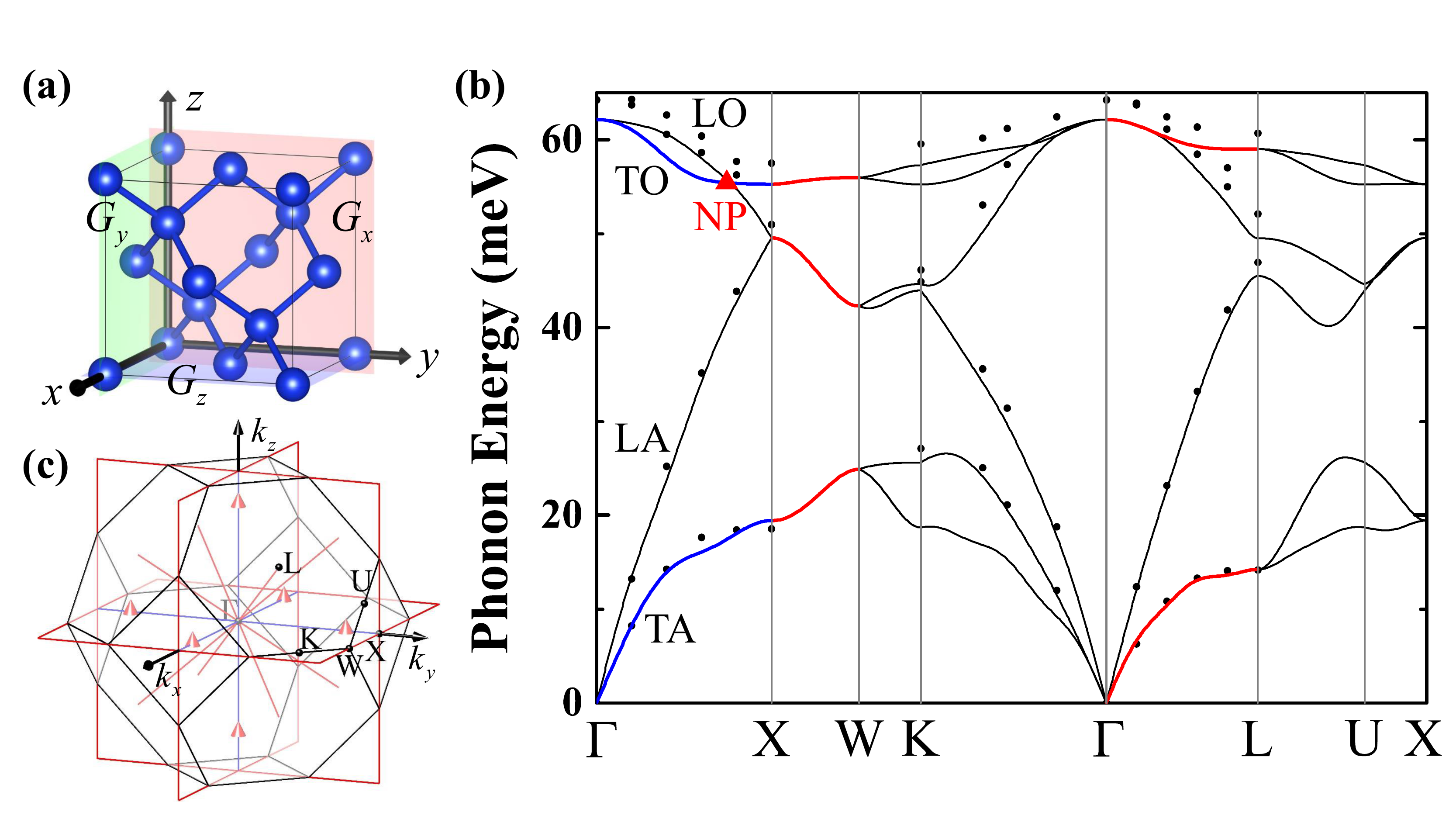}
\caption{\label{fig1}
(a) Diamond structure with glide mirror symmetries $G_{x,y,z}$. (b) Phonon dispersion of Si calculated by density functional theory (DFT, see methods in Supplemental Material (SM) Sec. S1 \cite{SM}) and measured by experiment (black dots) \cite{Dolling1963}. The red (blue) lines represent nodal lines with $\mathbb{Z}_2 = 1$ ($\mathbb{Z}_2 = 0$), and the red triangle denotes topological nexus phonon (NP). (c) Distribution of symmetry-enforced nodal lines and nexus phonons in the extended Brillouin zone.}
\end{figure}

\textit{Topological nodal line.}---Bulk Si has a diamond structure with space group $Fd\bar{3}m$, which has three glide-mirror, six mirror, and several (screw) rotation symmetries (Fig. \ref{fig1}(a)). The crystalline symmetry plays an essential role in determining band degeneracy and topology of phonons (Fig. \ref{fig1}(b)). As enforced by glide-mirror symmetry (see details in the Supplemental Material (SM) Sec. S2 \cite{SM}), the six phonon bands form three doubly degenerate nodal lines along XW, showing a 3D periodic cage-like nodal structure in $\mathbf{k}$ space [Fig. \ref{fig1}(c)]. We find that these nodal lines all have quantized Berry phase $\theta = \pi$. In analogy to electrons~\cite{Xiao2010}, the Berry phase for a given phonon band $n$ is defined as $\theta_{n} = \oint_C \mathbf{A}_{n\mathbf{k}} \cdot \text{d}\mathbf{k}$, where the Berry connection $\mathbf{A}_{n \mathbf{k}} = -i \langle \mathbf{u}_{n\mathbf{k}}| \nabla_{\mathbf{k}} \mathbf{u}_{n\mathbf{k}} \rangle$, $\mathbf{u}_{n\mathbf{k}}$ is the phonon eigenstate. The Berry phase of a nodal line $\theta = \theta_{n}$, where $n$ denotes one of bands forming the nodal line (usually the lower-frequency band) and $C$ is a closed loop encircling the nodal line. The Berry phase factor $\mathrm{exp}(i\theta)$ can only take quantized values of $\pm 1$ in the presence of $\mathcal{PT}$ symmetry~\cite{Kim2015PRL}. Since the TRS of phonon is hardly broken~\cite{LiuYZ2017PRB}, a $\mathbb{Z}_2$ topological classification is allowed for phonons of centrosymmetric materials: $\mathbb{Z}_2 = \mathrm{mod}(\theta/\pi, 2)$. The glide-mirror symmetry enforced nodal lines are topologically nontrivial ($\mathbb{Z}_2 = 1$). Their band degeneracy is movable but not removable under weak perturbations preserving $\mathcal{PT}$ symmetry, as supported by numerical calculations in SM Sec. S3~\cite{SM}.

\begin{figure}
\centering
\includegraphics[width=\linewidth]{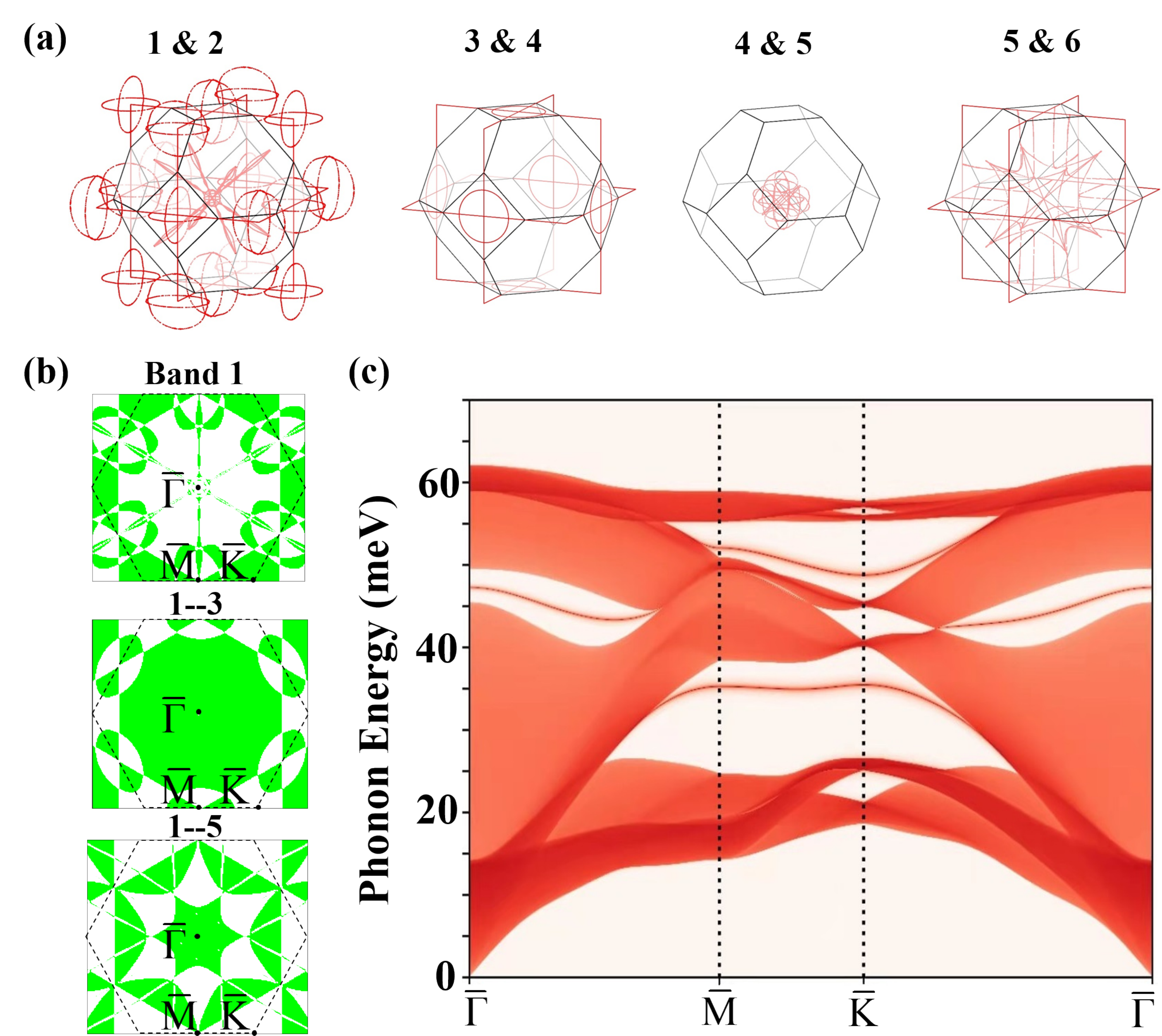}
\caption{\label{fig2}
(a) Distribution of topological nodal lines formed by bands $n$ and $n+1$ in the extended Brillouin zone (BZ). (b) Distribution of $\varphi_n(\mathbf{k}_\parallel)$ summed over different bands in the (111) surface BZ. Regions with $\varphi=\pi$ are colored green. (c) Local density of states (LDOS) of semi-infinite (111) surface calculated using the fixed boundary condition.}
\end{figure}

Moreover, mirror symmetry can protect nodal lines that are formed by bands with opposite mirror eigenvalues. Remarkably, {\it such kind of nodal lines are generally topologically nontrivial and ubiquitously exist} in the momentum space of Si (Fig. \ref{fig2}(a)). Linear order terms of $\mathbf{k}$ are usually allowed in the low-energy effective Hamiltonian of nodal lines, which gives $\mathbb{Z}_2 = 1$. In rare cases, the linear terms are forbidden by symmetry and the quadratic terms get dominating (e.g., for nodal lines along $\Gamma$X in SM Sec. S4 \cite{SM}), which changes $\mathbb{Z}_2$ to 0. In addition, the $\mathcal{PT}$ symmetry, mirror symmetries ensure another $\mathbb{Z}_2$ topological classification by defining the Berry phase of Wilson loop perpendicular to a 2D plane: $\varphi_{n}(\mathbf{k}_\parallel) = \int_{-\pi}^{\pi} \mathbf{A}_{n\mathbf{k}} \cdot \text{d}\mathbf{k}_\perp$, where $\mathbf{k}=\mathbf{k}_\parallel + \mathbf{k}_\perp$ with $\mathbf{k}_\parallel$ and $\mathbf{k}_\perp$ being parallel and perpendicular to the 2D plane, respectively. The so-called Zak's phase is quantized to 0 or $\pi$ for mirror-symmetric planes~\cite{Xiao2010}.

Figure \ref{fig2}(b) shows the distribution of $\varphi_n(\mathbf{k}_\parallel)$ summed over different bands for the (111) surface. The Zak's phase $\varphi(\mathbf{k}_\parallel)$ is constant in the non-degenerate region and changes abruptly by $\pi$ when varying $\mathbf{k}_\parallel$ across the projection of a topological nodal line, which is consistent with the distribution of the topological nodal lines shown in SM Sec. S5 \cite{SM}. According to the bulk-boundary correspondence \cite{Hasan2010RMP, Qi2011RMP}, a topological boundary state exists in the $\mathbf{k}_\parallel$ region of $\varphi = \pi$. This is evidenced by the surface-state calculation (Fig. \ref{fig2}(c)). Importantly, the drumhead-like topological surface states~\cite{Fang2016CPB} are separated from bulk states in some frequency windows, facilitating experimental observations. Note that for the (111) surface, the distribution of drumhead surface states is different for various surface structures, as explained by the fact that the Zak's phase is dependent on the gauge choice of bulk unit cell described in SM Sec. S6~\cite{SM}. Results of (001) and (110) surfaces are included in SM Sec. S7~\cite{SM}. The ubiquitous existence of topological nodal-lines in Si makes the topological surface states observable for arbitrary surface orientations.

\begin{figure}
\centering
\includegraphics[width=\linewidth]{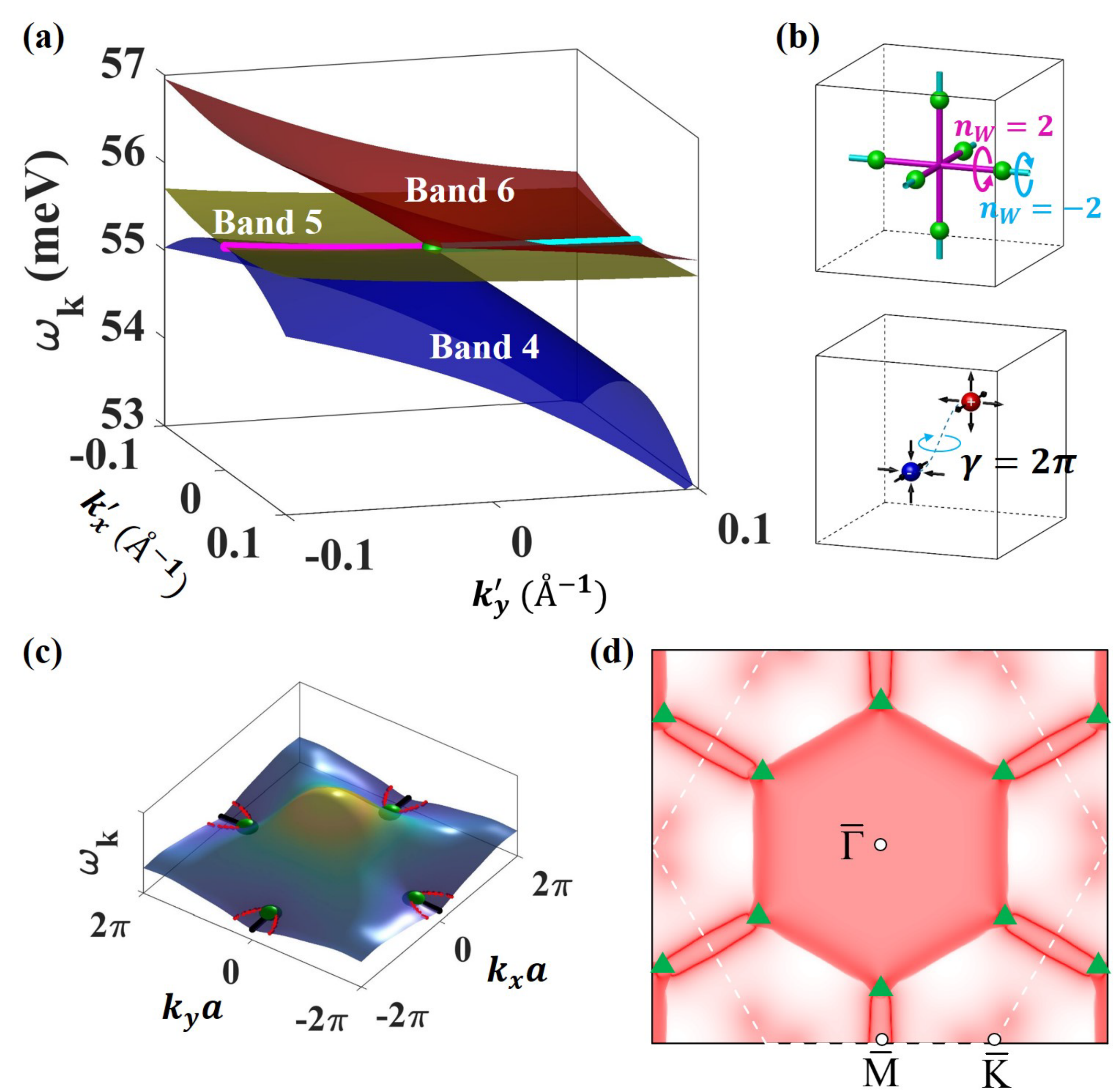}
\caption{\label{fig3}
(a) Schematic phonon dispersion near the triply degenerate point (TDP), formed by band crossing between a doubly degenerate band with a non-degenerate band along $\Gamma$X (i.e. $k_y$ direction). (b) Momentum-space nexus phonons with physical nodal strings in Si (upper panel), Dirac monopoles with invisible Dirac string (lower panel). The nodal string is singular as characterized by nonzero topological winding number  $n_W$. (c) Schematic of topological surface phonon bands of the (001) surface. The surface bands are pinned to the nodal strings. A constant-frequency contour (red lines) is shown for nexus phonons. (d) LDOS of Si(111) with frequency equal to that of nexus phonons ($\hbar\omega=55.2$ meV). The TDPs are denoted by green triangles.
}
\end{figure}

\textit{Topological nexus phonon.}---Nodal line is normally closed in the momentum space, thus also called nodal ring. Here we find a novel kind of composite nodal structure along $\Gamma$X in Si, which displays nodal strings terminating at exceptional points. As illustrated in Fig. \ref{fig3}(a), a nodal string formed by bands 4$\&$5 terminates at a triply degenerate point (TDP), where another nodal string formed by bands 5$\&$6 emerges. The TDP does not appear isolatedly as Weyl-like or Dirac-like nodal points, but connects with its inversion image via a line degeneracy (Fig. \ref{fig3}(b)). By symmetry there exist three pairs of TDPs in the system.

\begin{figure}
\centering
\includegraphics[width=\linewidth]{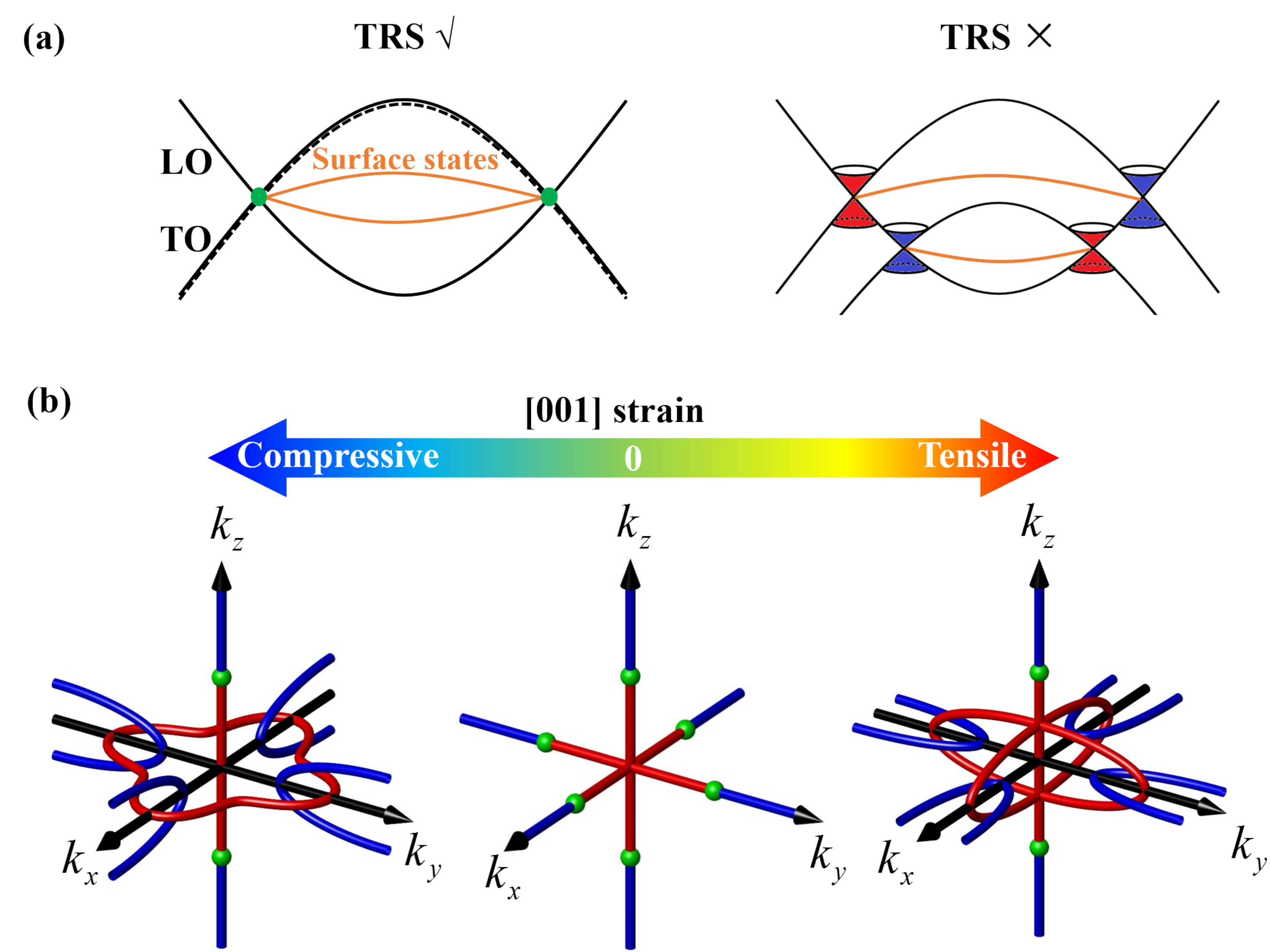}
\caption{\label{fig4}
(a) Nexus phonons with double Fermi-arc-like surface states (left panel) and Weyl phonons with Fermi-arc-like surface states formed by breaking TRS (right panel). (b) Schematic evolution from nexus phonons with nodal strings (middle panel) to topological Hopf nodal links under compressive or tensile strain along [001] direction.
}
\end{figure}

For understanding the topological nature, a three-band effective Hamiltonian model is derived near the TDP based on symmetry analysis shown in SM Sec. S8~\cite{SM}:
\begin{equation}\label{H0}
H_0 = \left(
\begin{array}{ccc}
C_1 k_y & 0 & t(k_x - k_z) \\
0 & C_1 k_y & t(k_x + k_z) \\
t^*(k_x - k_z) & t^*(k_x + k_z) & C_2 k_y
\end{array}
\right) + O(k^2),
\end{equation}
where the coefficients $C_1$ and $C_2$ are real ($C_1 > C_2$) whereas $t$ is imaginary, $\mathbf{k}$ is referenced to the TDP, and the line degeneracy is along $k_y$. For the doubly degenerate subspace with $k_y \ne 0$, the model can be simplified to two-band by L\"owdin partitioning \cite{Cardona2005}: $\bar H_\text{D} = A_\mathbf{k} \sigma_x + B_\mathbf{k} \sigma_z + C_\mathbf{k} $, where $A_\mathbf{k}=  \frac{(k^2_x - k^2_z)|t|^2}{(C_1 - C_2)k_y}$, $B_\mathbf{k}=  \frac{-2k_xk_z|t|^2}{(C_1 - C_2)k_y}$, $C_\mathbf{k} = C_1 k_y + \frac{(k^2_x + k^2_z)|t|^2}{(C_1 - C_2)k_y}$, $\sigma_{x,y,z}$ are Pauli matrices. By changing the energy reference, the term $C_\mathbf{k}$ can be removed without affecting topological properties. Then the model Hamiltonian $H_\text{D} = \bar H_\text{D} -C_{\mathbf{k}}$ obeys a chiral symmetry $\{\sigma_y, H_{\text{D}}\} =0$, and a topological winding number $n_W = $Tr$\oint_C \frac{\text{d}l}{4\pi i} \left[ \sigma_y H^{-1}_{\text{D}} \partial_l H_{\text{D}} \right]$ is defined for an infinitesimal loop $C$ encircling the line degeneracy~\cite{Heikkil2015NJP}. Our calculations indicate that the nodal strings along $\Gamma$X have $n_W = -2$sgn$(k_y)$, which changes sign across the TDP (Fig. \ref{fig3}(b)).

The nodal strings with nonzero  $n_W$ are singular and cannot be adiabatically gapped out. According to topological theory, such a singular string must terminate at a monopole. The monopole emanating singular string is named nexus in relativistic theory~\cite{Volovik2003}. In analogy to Dirac monopole, the nexus is always attached with ``Dirac string''. Differently, the string is invisible for Dirac monopole but physical for the nexus (Fig. \ref{fig3}(b)). Real space nexus has been found in $^3$He-A and chiral superconductor~\cite{Volovik2003}, and momentum space nexus of electrons in topological materials~\cite{Heikkil2015NJP}. Momentum space nexus of phonons is discovered here in Si. The nexus is composed of a pair of TDPs connected by a nodal string and have nonzero Euler monopole charges in momentum space as required by $\mathcal{PT}$ symmetry with more detail in SM Sec. S9 \cite{Lenggenhager2021, SM}.

The nexus phonons are able to show rich topological features. Based on the effective Hamiltonian (Eq. \eqref{H0}), we derived a tight binding model for the three-band system and systematically studied its topological properties in SM Sec. S8~\cite{SM}. The topological nontrivial nature of nexus phonons is embodied by bulk-boundary correspondence. For the (001) surface, we find that the Zak's phase of the upper band unusually has $\varphi_n(\mathbf{k}_\parallel) \equiv \pi$. As the nodal strings have $\mathbb{Z}_2 = 0$, the quantized $\varphi_n(\mathbf{k}_\parallel)$ does not change abruptly across them. The nonzero constant $\varphi_n(\mathbf{k}_\parallel)$ results in topological boundary states throughout the whole surface Brillouin zone. The topological surface bands have to merge into bulk bands via nodal strings. They are nearly flat near nodal strings and get dispersive elsewhere, showing a ``hat''-like shape (Fig \ref{fig3}(c)). In the constant-frequency contour, every TDP is connected by a pair of Fermi-arc-like surface states whose group velocities are opposite near the TDP. This is consistent with the picture that the TDP is a composite of two opposite monopoles. By {\it ab initio} phonon calculations, we find that the double Fermi-arc-like surface states of nexus phonons can be clearly visualized for the Si(111) surface (Fig. \ref{fig3}(d)).

The composite nodal structure of nexus-string with intricate three-band entanglement cannot be fully eliminated by weak perturbations. Instead, it will transform into versatile topological states under symmetry-breaking fields. The physical picture is clarified via the effective Hamiltonian $H_\text{D}$. Under TRS-breaking (e.g. magnetic or Coriolis) field along $\Gamma$X, a perturbation term is introduced into $H_\text{D}$: $\Delta H_\text{T} = \delta_\text{T} \sigma_y$ as derived in SM Sec. S8~\cite{SM}. This breaks the line degeneracy and splits each TDP into a pair of Weyl nodes  with opposite topological charges, leading to a topological transition from nexus to Weyl phonons (Fig. \ref{fig4}(a)). The double Fermi-arc-like surface states get disconnected and evolve into two separated arcs.

Under a [001] strain field that reduces $C_4$ rotation symmetry into $C_2$ whereas preserves $M_x$ and $M_z$ mirror symmetries, a perturbation $\Delta H_\mathrm{S} = \delta_\mathrm{S} \sigma_x$ is added into $H_\text{D}$ as derived in SM Sec. S8~\cite{SM}. The TDP is destroyed, but the line degeneracy not (Fig. \ref{fig4}(b)). The latter is shifted away from $\Gamma$X and changes into two mirror-symmetric copies. The newly generated nodal lines have $n_W = \pm 1$ and $Z_2 = 1$. Their momentum space location is determined by $A_\mathbf{k} + \delta_\mathrm{S} = 0$ and $B_\mathbf{k}= 0$. Near the original nexus point, a nodal line is located in the $M_x$ plane on one side (e.g. $k_y >0$), another one in the $M_z$ plane on the opposite side. The situation is reversed when varying the sign of $\delta_\mathrm{S}$. Remarkably, the nodal lines always interlock with each other (Fig. \ref{fig4}(b)), which cannot be adiabatically unlocked as protected by topology. The  interlocking nodal loop is named topological Hopf nodal link~\cite{Chang2017PRL, XieY2019PRB}. This novel topological identity is closely related to non-Abelian quaterion charge~\cite{Wu2019Science}, which is rarely studied for phonons. Due to the nonzero $n_W$ and $Z_2$, the Hopf nodal link is expected to give flat drumhead-like surface phonon modes with large density of states, which might be important to surface-phonon related physics, such as low-dimensional superconductivity.

\begin{table}
\centering
\caption{\label{tab1} List of centrosymmetric space groups that give symmetry-enforced nodal point, line, or plane of phonons.}
\begin{tabular}{p{2cm}|p{6cm}}\hline\hline
$\quad$ & Space group No. \\\hline
Nodal point & 52, 54, 56, 124, 126, 128, 130, 133, 135, 137, 138, 163, 165, 167, 176, 192, 193, 194, 201, 206, 222--224, 226--228, 230 \\
Topological nodal line & 13--15, 48--50, 52--54, 56, 58, 60, 64, 66--68, 70, 72--74, 85, 86, 88, 124--126, 128, 130--138, 140--142, 163, 165, 167, 176, 192--194, 201, 203, 206, 222--224, 226--228, 230 \\
Nodal plane & 11, 14, 51--64, 127, 129, 130, 135--138, 176, 193, 194, 205 \\\hline
\end{tabular}
\end{table}

In addition to Si, we also find a few other candidate materials of nexus phonons, including diamond-structure materials C, Ge, and Sn, and zinc-blend-structure materials GaAs, InAs, InSb, \textit{etc}. Their nexus phonons are formed by band crossing between longitudinal optical branch and doubly degenerate transverse optical branch along $\Gamma$X, as found in Si.

Experimentally, the first microscopic surface phonon measurement was performed on Si(111) by Ibach using electron energy loss spectroscopy (EELS) one half century ago~\cite{Ibach1971PRL}. An optical surface phonon mode was discovered, whose physical origin remains elusive despite intensive study~\cite{DiNardo1986PRB, Harten1988PRB, Pennino1989PRB, Alerhand1985PRL, Miglio1989PRL, Ancilotto1990PRL}. Noticeably, this mysterious surface phonon is within the frequency region of 55-60 meV, coinciding with that of nexus phonons ($\sim$55 meV from theory). Similar surface phonons were detected on Si(001)~\cite{Takagi1999PRB} and Si(110)~\cite{Matsushita2014} as well, supporting the topological origin. Moreover, many other surface phonon modes with different frequencies were also observed in Si, consistent with the ubiquitous existence of topological phonons. In-depth experimental studies are required to study the topological nature of phonons in this important material, for instance, by inelastic helium atom scattering or angle-resolved EELS.

\textit{Generalization to other centrosymmetric materials.}---As learned from Si, the nonsymmorphic symmetry plays an important role in giving band degeneracy with nontrivial topology. In 3D space there are 157 nonsymmorphic space groups (SGs) and 68 of them are centrosymmetric. The nonsymmorphic SGs can enforce degeneracy of points, lines, or even planes at Brillouin zone boundary. A full classification of symmetry-enforced nodal structures for the centrosymmetric materials is presented in Tab. \ref{tab1}. For a given SG, we first check little groups of high-symmetry $\mathbf{k}$ points, lines or planes, and focus on those having 2D or higher dimensional irreducible representations. The analysis reveals that 27 SGs have 3-, 4-, or 6-fold point degeneracies at high-symmetry points, 56 SGs support topological nodal lines with $\mathbb{Z}_2 = 1$, and 27 SGs have nodal planes. Note that most nodal lines have Berry phase of $\pi$, except those protected by $C_{4v}$ or $C_{6v}$ symmetries. For all these SGs, we list the $k$-space positions of symmetry-enforced nodal structures explicitly and present phonon dispersions of candidate materials (adapted from Ref.~\onlinecite{Togo2015ScriptaMaterialia}) in SM Sec. S10 ~\cite{SM}. A full discussion of topological properties will be presented elsewhere.

In summary, we find that topological phonon states ubiquitously exist in solids, offering new opportunities to explore topological and Berry-phase physics of phonons in realistic materials. Silicon can serve as an ideal model material for the research,  not only because a novel kind of nexus phonon is predicted together with versatile topological states tuned by external fields, but also because the material is of key importance to semiconductor industry whereas the influence of topological surface phonons on device performance is largely unknown but might be critical especially to developing next-generation nanochips.

\begin{acknowledgments}
We thank Atsushi Togo for sharing the raw data of phonon database and Dr. Chong Wang for sharing a mathematica package to construct $k\cdot p$ models for crystals. This work was supported by the Basic Science Center Project of NSFC (Grant No. 51788104), the Ministry of Science and Technology of China (Grants No. 2016YFA0301001, No. 2018YFA0307100, and No. 2018YFA0305603), the National Natural Science Foundation of China (Grants No. 11874035, No. 12074091, and No. 11334006), and the Beijing Advanced Innovation Center for Future Chip. Y.X. acknowledges support from the National Science Fund for Distinguished Young Scholars (Grant No. 12025405). The calculations were done on the ``Tianhe-2'' system of the National Supercomputer Center in Guangzhou. Y.L. and N.Z. contributed equally to this work.
\end{acknowledgments}

%

\end{document}